\documentclass[twocolumn]{aastex63}
\usepackage{amssymb}

\newcommand{\msun}{{\rm M}_\odot}
\newcommand{\kms}{{\rm km\,s^{-1}}}

\received{}
\revised{}
\accepted{}
\submitjournal{ApJ}
\shorttitle{}
\shortauthors{}
\graphicspath{{./}{figures/}}
\begin{document}
\title{Revealing a Centrally Condensed Structure in OMC-3/MMS 3  \\with ALMA High Resolution Observations}
\author{Kaho Morii}
\affiliation{Department of Astronomy, Graduate School of Science, The University of Tokyo, 7-3-1 Hongo, Bunkyo-ku, Tokyo 113-0033, Japan email: kaho.morii@grad.nao.ac.jp}
\affiliation{Division of Science, National Astronomical Observatory of Japan, 2-21-1 Osawa, Mitaka, Tokyo 181-8588, Japan}
\affiliation{Department of Earth and Planetary Sciences, Faculty of Sciences, Kyushu University, Fukuoka 819-0395, Japan}

\author{Satoko Takahashi }
\affiliation{ Joint ALMA Observatory, Alonso de Córdova 3107, Vitacura, Santiago, Chile}
\affiliation{NAOJ Chile, National Astronomical Observatory of Japan, Alonso de Córdova 3788, Office 61B, Vitacura, Santiago, Chile, 7630492}
\affiliation{Department of Astronomical Science, School of Physical Sciences, The Graduate University for Advanced Studies, SOKENDAI, Mitaka, Tokyo 181-8588, Japan}
\author{Masahiro N. Machida}
\affiliation{Department of Earth and Planetary Sciences, Faculty of Sciences, Kyushu University, Fukuoka 819-0395, Japan}

\begin{abstract}
Using the Atacama Large Millimeter/submillimeter Array (ALMA), we investigated a peculiar millimeter source MMS 3 located in the Orion Molecular Cloud 3 (OMC-3) region in the 1.3 mm continuum, CO ($J$=2--1), SiO ($J$=5--4), C$^{18}$O ($J$=2--1), $\rm{N_2D^+}$ ($J$=3--2), and DCN ($J$=3--2) emissions. 
With the ALMA high angular resolution ($\sim$0$\farcs2)$, we detected a very compact and highly centrally condensed continuum emission with a size of $0\farcs45 \times 0\farcs32$ (P.A.=0.22$^\circ$).
The peak position coincides with the locations of previously reported $Spitzer$/IRAC and X-ray sources within their positional uncertainties.
We also detected an envelope with a diameter of $\sim$6800\,au (P.A.=75$^\circ$) in the C$^{18}$O ($J$=2--1) emission. 
Moreover, a bipolar outflow was detected in the CO ($J$=2--1) emission for the first time. The outflow elongates roughly perpendicular to the long axis of the envelope detected in the C$^{18}$O ($J$=2--1) emission. Compact high-velocity CO gas in the (red-shifted) velocity range of 22--30 km\,s$^{-1}$, presumably tracing a jet, was detected near the 1.3 mm continuum peak.
A compact and faint red-shifted SiO emission was marginally detected on the CO outflow lobe. 
The physical quantities of the outflow in MMS 3 are relatively smaller than those in other sources in the OMC-3 region.
The centrally condensed object associated with the near-infrared and X-ray sources, the flattened envelope, and the faint outflow indicate that MMS 3 harbors a low mass protostar with an age of $\sim$10$^3$ yr.
\end{abstract}

\section{Introduction}
\label{sec:intro}
It is crucially important to identify how and when stars form in their natal clouds to understand the whole picture of star formation.
However, it is difficult to identify newborn stars in gravitationally collapsing cloud cores because the growth timescale of protostars is very short.
In a gravitationally collapsing cloud, the first hydrostatic core (or the first Larson core, hereafter the first core, \citealt{larson1969,masunaga00}) forms  before protostar formation. 
After protostar formation, the first core (remnant) remains around the protostar and becomes a rotationally supported disk \citep{Bate1998,Machida2010}.
The first core remnant or  rotationally supported disk drives an outflow that determines the star formation efficiency \citep{Tomisaka2002, Machida2012}.
Thus, the first core and its remnant play a key role in the early star formation phase \citep{Bate1998,masunaga00,Machida2010}.
However, the timescale for the first core is as short as $\sim$10$^2$--$10^3$\,yr \citep{larson1969,masunaga00,Saigo2006}. 
The protostar phase, during which parcels of gas continue to accrete onto the protostar or circumstellar disk from the infalling envelope, lasts for $\sim$10$^5$\,yr. 
Thus, the detection rate for the first core is 1/100--1/1000 (($10^2$--$10^3$\,yr)/$10^5$\,yr).  
Although the detection rate for  very young protostars is also low, these objects provide  useful information for understanding the star formation process.

The Orion Molecular Cloud 3 (OMC-3) star-forming region is located $d=392$\,pc from the Sun \citep{Tobin2020} and is one of the best sites for investigating the early phase of star formation. 
In OMC-3, there are both prestellar and protostellar sources MMS 1--10, which are identified in  \citet{chini1997}. 
\citet{takahashi2013} reported that cores  are formed by fragmentation of filaments. 
As seen in \citet{takahashi2013}, the main filament in northern OMC-3 extends from the northwest to the southeast (P.A. =  135$^\circ$).
The protostellar sources in this region contain Class 0 and I protostars \citep[e.g.,][]{chini1997,Nielbock2003,Takahashi08,Furlan2016}. 
In our previous studies, we reported the detailed properties of the Class 0 protostellar sources MMS 5 and MMS 6 and associated compact and collimated outflows, which were observed by the Submillimeter Array (SMA) and ALMA \citep{Takahashi08,Takahashi09,TakahashiandHo2012,Takahashi2012,takahashi2013,Takahashi2019,Matsushita2019}.

In this paper, we focus on another millimeter source, MMS 3, that is identified as SMM 4 in \citet{takahashi2013}, located around the center of the OMC-3 filament (for details, see Figure~3 of \citealt{takahashi2013}).
The average number density of MMS 3 estimated from the SMA 0.85 mm dust continuum observation is $n= (4\pm 2) \times 10^6$\,cm$^{-3}$,  which is one order of magnitude lower than the other very young protostellar sources in OMC-3 \citep{takahashi2013}. 
The peak flux of MMS 3 is 4--20 times weaker than the other protostellar sources in this region. 
In addition, to date, neither outflow nor jet has been observed in MMS 3 \citep[][]{Reipurth99,Aso00,Williams03,Takahashi08,Tanabe19,Feddersen20}.
Moreover, no 9 mm continuum emission has been detected around MMS 3 in the recent VLA observations \citep{Tobin2020}.
Thus, it appears that MMS 3 is in the prestellar phase before gravitational contraction has begun. 
Nevertheless, X-ray emission was detected by the $Chandra\ X-Ray\ Observatory$ (TKH 10, \citealt{Tsuboi2001}).
Furthermore, an infrared source was also detected in MMS 3 in both the $Spitzer$, with wavelengths longer than the IRAC 4.5 $\mu$m bands, and the $Herschel/PACS/SPIRE$ bands (HOPS 91 in \citealt{Furlan2016}, Megeath 2442 in \citealt{Megeath2012}). 
Thus, although we could not detect a central condensation in MMS 3 in the SMA observation, a protostar has already been  born within MMS 3.
In this study, we investigate this peculiar object MMS 3 with ALMA observations to elucidate the early phase of star formation.

This paper is structured as follows.
In \S \ref{sec:obs}, we explain the observation method and data reduction process.
The results are shown in \S \ref{sec:results}. 
We discuss the evolutionary stage of MMS 3 in \S \ref{sec:discussion} and summarize our results in \S \ref{sec:summary}.

\section{Observations}
\label{sec:obs}
The observations were made in Cycle 3 (2015.1.00341.S, PI. S. Takahashi), 
2016 June 30, July 12, 17, and 19  (Atacama Compact 7m-array, hereafter the ACA),
2016 January 29 (12 m array in the compact configuration C36-1, 
hereafter ALMA low resolution), and 2016 September 18 and 19
(12 m array in the extended configuration C36-4, 
hereafter ALMA high resolution) using the 1.3 mm band (Band 6). 
The phase center was set at 
R.A. (J2000) = 5$^h$35$^m$18$\fs$980, 
Dec (J2000) = $-$05$^{\circ}$00$^{\prime}$51$\farcs$630. 
The observation parameters are listed in Table 1. 
The total on-source time was 16 minutes for the ACA, 
4 minutes for the  ALMA low resolution, 
and 16 minutes for the ALMA high resolution.
The datasets cover projected baselines 
between 5.5 k$\lambda$ and 365 k$\lambda$ (ACA), 
8.6 k$\lambda$ and 240 k$\lambda$ (ALMA low  resolution), 
and 9.0 k$\lambda$ and 2400 k$\lambda$ (ALMA high  resolution).
These were insensitive to structures extended more than 21$\arcsec$ for the ACA, and 13$\arcsec$ for the ALMA low and high resolution 
at the 10$\%$ level of the total flux density (ALMA Cycle 3 Technical handbook). 
Our spectral setups include CO ($J$ = 2--1), SiO ($J$ = 5--4), 
C$^{18}$O ($J$ = 2--1), N$_2$D$^{+}$ ($J$ = 3--2), 
and DCN ($J$ = 3--2) emissions. 
These were observed with velocity resolutions of 
0.37 km s$^{-1}$, 0.39 km s$^{-1}$, 0.048 km s$^{-1}$, 
0.046 km s$^{-1}$, and 0.39 km s$^{-1}$, respectively. 
Line-free channels corresponding to effective bandwidths of 695 MHz (ACA), 800 MHz (ALMA low resolution), 
and 782 MHz (ALMA high resolution) were allocated 
for imaging the continuum emission. 
Calibration of the raw visibility data was performed by 
the ALMA observatory with the standard calibration method 
using the Common Astronomy Software Application 
(CASA; \citealt{McMullin07}) 
versions 4.6.0 for ACA and ALMA low resolution datasets and 4.7.0  for ALMA high resolution datasets.

After calibration, clean images were made using the CASA task ``clean''. 
Natural weightings were used for the final images. 
Velocity widths of 0.1 km s$^{-1}$ (C$^{18}$O and N$_2$D$^{+}$), 0.5 km s$^{-1}$ (DCN), and 1.0 km s$^{-1}$ (CO and SiO) were used to produce the image cubes.
Data combination from the two different arrays were performed in the image base 
for the 1.3 mm continuum, C$^{18}$O, N$_2$D$^+$, and DCN
using the CASA task ``feather'', 
and in the $uv$ base for CO using the CASA task ``concat''.
An image produced from the ALMA low resolution datasets was only presented for the SiO emission\footnote{ No data combination was performed since there was no total flux difference in the image between the ACA and ALMA low resolution data sets.}.
No primary beam correction has been applied for the presented images in this paper, while primary beam corrected CO and C$^{18}$O flux values were used for deriving physical parameters. 
The resulting synthesized beam sizes and 1$\sigma$ rms noise 
levels are listed in Table~\ref{tab:impara}.
 
\begin{deluxetable*}{lccc}
\tabletypesize{\scriptsize}
\tablecaption{ Observation Parameters}
\tablewidth{0pt}
\tablehead{
\colhead{Parameters} & \colhead{ACA} & \colhead{ ALMA low resolution} & \colhead{ALMA high resolution} }
\startdata
Observing date (YYYY-MM-DD)		&	2016-06-30 -07-12, -17 and -19	& 2016-01-29 & 	2016-09-18 and -19\\
Number of antennas				&	10         	& 40 			& 48			\\
Primary beam size (arcsec)      &	46	        & 27            & 27		    	\\
PWV (mm)						&	--	    & 2.6 	    & $\sim$2.2 and $\sim$1.1  	    \\
Phase stability rms (degree)\tablenotemark{a} & $\sim$10   & 	$\sim$13	    &	$\sim$21 and $\sim$52				\\
Bandpass calibrators  		    & J0538-4405 and J0522-3627	&  J0522-3627           & J0510+1800		\\
Flux calibrators     			&J0522-3627                 & J0522-3627 & J0522-3627 and	J0510+1800	\\
Phase calibrators\tablenotemark{b}   &  J0607-0834 and J0542-0913    &  J0541-0541          &  J0607-0834        \\
Total continuum bandwidth; USB+LSB (MHz)& 695   & 800          & 782	        \\
Projected baseline ranges (k$\lambda$)  & 5.5--365 & 8.6--240	    & 9.0--2400	    \\
Maximum recoverable size (arcsec)\tablenotemark{c}	& 21 	& 13 & 13	        \\
Total on-source time (minutes)      	&16     & 4			& 16
\enddata
\tablenotetext{a}{Antenna-based phase differences for the bandpass calibrators.}
\tablenotetext{b}{The phase calibrator was observed every 8 minutes.}
\tablenotetext{c}{Our observations were insensitive to emissions more extended than this size scale at the 10\% level of the total flux density (ALMA Cycle 3 Technical Handbook).}
\label{tab:obspara}
\end{deluxetable*}

\begin{deluxetable*}{lcccc}
\tabletypesize{\scriptsize}
\tablecaption{Image Parameters}
\tablewidth{0pt}
\tablehead{
\colhead{Image}  & \colhead{Synthesized beam size (P.A.)} & \colhead{RMS noise level} & \colhead{Velocity width} & \colhead{Figure} \\ \colhead{} & \colhead{arcsec $\times$ arcsec  (deg)} & \colhead{} & \colhead{\,km\,s$^{-1}$} &\colhead{}}
\startdata
ACA Continuum & 8.2 $\times$ 4.8 (-90) & 4.4\,[mJy\,beam$^{-1}$] & -- & \ref{fig:continuum}(a) \\
ACA + ALMA low resolution continuum\tablenotemark{a} & 1.8 $\times$ 1.0 (-74) & 0.46\,[mJy\,beam$^{-1}$] & -- & \ref{fig:continuum}(b) \ref{fig:C18ON2D+} \ref{fig:C18ODCN}\\
ALMA high resolution continuum &0.22 $\times$ 0.18 (-22) & 0.093\,[mJy\,beam$^{-1}$] & -- & \ref{fig:continuum}(c)\\
ACA + ALMA low resolution C$^{18}$O($J$=2--1) \tablenotemark{a}& 1.8 $\times$ 1.0 (-73) & 16\,[mJy\,beam$^{-1}$] & 0.3 & \ref{fig:c18ochan}\\
ACA + ALMA low resolution C$^{18}$O($J$=2--1) \tablenotemark{a}& 1.8 $\times$ 1.0 (-73) & 29\,[mJy\,beam$^{-1}$\,km\,s$^{-1}$] & 4.4 &\ref{fig:C18ON2D+} \ref{fig:C18ODCN} \ref{fig:COSiO}\\
ACA + ALMA low resolution N$_2$D$^+$($J$=3--2) \tablenotemark{a} & 1.8 $\times$ 1.0 (-74)  & 34 \,[mJy\,beam$^{-1}$\,km\,s$^{-1}$] & 4.4 &\ref{fig:C18ON2D+}\\
ACA + ALMA low resolution DCN ($J$=3--2) \tablenotemark{a} &1.9 $\times$ 1.1 (-74)  & 23\,[mJy\,beam$^{-1}$\,km\,s$^{-1}$] & 9.5 &\ref{fig:C18ODCN}\\
ALMA low resolution SiO($J$=5--4) & 1.9 $\times$ 1.1 (-74) & 2.9\,[mJy\,beam$^{-1}$\,km\,s$^{-1}$] & 3.0 &\ref{fig:COSiO}\\
ACA + ALMA low resolution CO ($J$=2--1) \tablenotemark{b}& 1.7 $\times$ 1.0 (-74)  & 150, 200, 28\,[mJy\,beam$^{-1}$\,km\,s$^{-1}$]\tablenotemark{c} & 16, 22, 8.0\tablenotemark{c} &\ref{fig:COSiO}\\
\enddata
\tablenotetext{a}{Data combination was done in the image base using the CASA task ``feather''. }
\tablenotetext{b}{Data combination was done in the UV visibility base using the CASA task ``concat''.  }
\tablenotetext{c}{The different rms noise levels are due to the fact that the integrated intensity maps used different velocity ranges. A detailed description is given in the caption of Figure~\ref{fig:COSiO}}
\label{tab:impara}
\end{deluxetable*}

\section{Results} \label{sec:results}
\subsection{\rm Dust Continuum Emission} \label{subsec:continuum}
\begin{figure*}
\epsscale{1.2}
\plotone{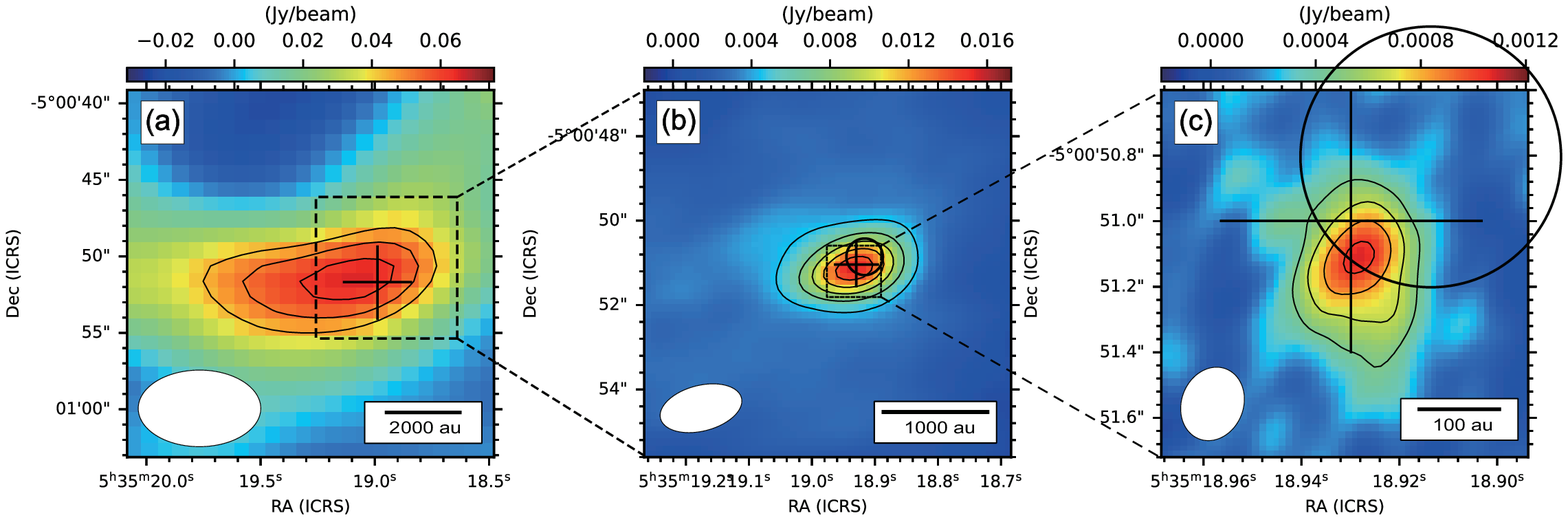}
\caption{
Dust continuum emission (color and black contours) obtained in the 1.3 mm band from (a) ACA 7 m array, (b) ACA + ALMA low resolution, and (c) ALMA high resolution.
The black contours show 10, 12, and 14$\sigma$ (1$\sigma$ = 4.4 mJy beam$^{-1}$) for panel (a), 10, 15, 20, 25, and 30$\sigma$ (1$\sigma$ = 0.46 mJy beam$^{-1}$) for panel (b), and 5, 7, 9, and 11$\sigma$ (1$\sigma$ = 0.093 mJy beam$^{-1}$) for panel (c).
The white ellipse in the bottom left corner represents the synthesized beam size in each image.
The black line in the bottom right corner of each panel indicates the spatial scale.
The cross in panel (a) corresponds to the positions of MMS 3 identified by the SMA 850 $\mu$m continuum emission \citep{takahashi2013}.
The circles and crosses in panels (b) and (c) correspond to the position of the infrared \citep[HOPS 91,][]{Furlan2016}) and X-ray sources \citep{Tsuboi2001}, respectively. The sizes of the circles and crosses indicate their positional uncertainties. }
\label{fig:continuum} 
\end{figure*}

Figure~\ref{fig:continuum} shows the 1.3 mm dust continuum emission. 
In Figure~\ref{fig:continuum}{\it a}, 
we can confirm a filamentary structure of the dust 
distribution in the east--west direction.
Using two-dimensional (2D) Gaussian fitting, 
the size of the filamentary structure is 
estimated to be $(20\arcsec \pm 1\farcs0) \times (7\farcs1 \pm 0\farcs30)$, 
which corresponds to $\sim$7800\,au $\times$ 2800\,au 
in the linear size scale with the position angle (P.A.) of $\sim$98$^\circ$. 
The peak position of the dust continuum emission is 
R.A. =  $05^h35^m19\fs185$ and 
Dec = $-05^{\circ}00^{\prime}51\farcs598$, 
which is shifted 4$\arcsec$ from the peak position 
obtained from the previous SMA observations 
in the 850 $\mu$m band \citep{takahashi2013}. 
We detected a more extended structure in the ACA observations (Figure~\ref{fig:continuum}(a)). This is because the maximum recoverable size of the ACA observations is 1.6 times larger than that in previous SMA observations.
The ACA observations also achieved $\sim$2.3 times better 
sensitivity compared with previous SMA observations.\footnote{In order to directly compare the sensitivity between 
the ALMA 1.3 mm data and the SMA 0.85 mm data, 
the beam surface area difference and 
the spectral index between the two wavelengths 
($\beta$=1.6 from \citealt{Johnstone_1999}) were taken into account.}
These improved sensitivities allow the detection 
of more extended and faint emissions in the ALMA images.
Within the extended structure, for the first time, 
we detected a compact and centrally condensed millimeter source 
with the ALMA observations (Figures~\ref{fig:continuum}(b) and (c)). 
The peak position of the 1.3\,mm continuum emission in Figure~\ref{fig:continuum}(c) is R.A. = $05^h35^m18^s.929$ 
and Dec = $-05^{\circ}00^{\prime}51\farcs 133$. 
The 1.3 mm continuum emission peak coincides with the previously reported infrared (detected wavelength longer than 4.5\,$\mu$m, \citealt{Furlan2016}) 
and X-ray  \citep{Tsuboi2001} sources within their positional errors. 
Thus, both the infrared and X-ray sources are associated 
with the centrally condensed structure. 
Using 2D Gaussian fitting, the compact 1.3 mm continuum structure 
in Figure~\ref{fig:continuum}(b) was measured to be 
($1''.8 \pm 0''.3) \times (1''.1 \pm 0''.13)$, corresponding 
to a linear size of  $700 \times 430$\,au.
The size of the very compact source detected 
with the ALMA high resolution 
(Figure~\ref{fig:continuum}(c)) is  
(0$''.45 \pm 0''.031) \times (0''.32 \pm 0''.023)$.  
This corresponds to a linear size of $\sim$180 $\times$ 130\,au.
The deconvolved size of MMS 3 reported by \citet{Tobin2020} 
is $0''.26 \times 0''.20$. Thus,  our result is in agreement 
with \citet{Tobin2020} within a factor of 2.

Using the ratio between the major and minor axes of the 1.3\,mm continuum emission, we can estimate the inclination angle of the disk $i$ with respect to the plane of the sky.
The angle calculated from the ALMA high resolution image (Figure~\ref{fig:continuum}c) is $i=44^\circ$.
On the other hand, the inclination angle derived from \citet{Tobin2020} is $i=40^\circ$.
Thus, our results are consistent with \citet{Tobin2020} within  $\sim$5$^\circ$.

Using 2D Gaussian fitting, we also obtained the total flux
and peak flux for each image in Figure~\ref{fig:continuum}, summarized in Table~\ref{tab:contdata}.
A structure comparable to the beam size was detected 
in the ALMA high resolution (Figure~\ref{fig:continuum} 
and Table~\ref{tab:contdata}).
Assuming optically thin dust thermal emission and 
an isothermal dust temperature, we estimate the dust mass 
with the total 1.3 mm flux, $F_{\rm 1.3mm}$, using 

\begin{equation}
M_{\rm dust}=\frac{F_{\rm 1.3mm}d^2}{\kappa_{\rm 1.3mm}B_{\rm 1.3mm}(T_{\rm \rm dust})},
\end{equation}
where $d$, $\kappa_{\rm 1.3mm}$, and  $B_{\rm 1.3mm}(T_{\rm dust})$ 
are the distance to the source (392\,pc, \citealt{Tobin2020}), the 
absorption coefficient for the dust per unit mass, 
and the Planck function as a function of the dust temperature 
$T_{\rm dust}$, respectively. 
We adopt $\kappa_{\rm 1.3mm} = 0.9$\,cm$^2$\,g$^{-1}$ 
from the dust coagulation model of the MRN \citep[][]{Mathis77}
distribution with thin ice mantles computed by \citet{OssenkopfHenning1994}. 
Assuming a dust temperature of $T_{\rm dust} =20$ and 40\,K, we obtain the dust mass, 
column density, and number density values shown in Table~\ref{tab:mass}. 
The gas masses estimated from Figure~\ref{fig:continuum}({\it c}) are (1.6 $\pm$ 0.1) $\times 10^{-2}$ M$_\odot$ for $T_{\rm dust}=$ 20\,K and (7.0 $\pm$ 0.4) $\times 10^{-3}$ M$_\odot$ for $T_{\rm dust}=$ 40\,K, in which a gas-to-dust mass ratio of 100 is assumed.

\begin{deluxetable*}{lccc}
\label{tab:contdata}
\tabletypesize{\scriptsize}
\tablecaption{The 1.3 mm Continuum Source Parameters}
\tablewidth{0pt}
\tablehead{
\colhead{Data} & \colhead{Total flux (mJy)} & \colhead{Peak flux (mJy\,beam$^{-1}$)} & \colhead{Deconvolved size  (arcsec $\times$ arcsec,\,deg)}}
\startdata
ACA & 310 $\pm$ 12 & 64 $\pm$ 2.1  & 20 $\pm 1.0 \times 7.1 \pm 0.30, 98 \pm 1.2$ \\
ACA + ALMA low resolution & 31 $\pm$ 2.6 &15 $\pm$ 0.87& 1.8 $\pm$ 0.3 $\times$ 1.1$ \pm$ 0.13, 100 $\pm$ 15\\
ALMA high resolution & 4.6 $\pm$0.28 & 1.0 $\pm$ 0.050 & 0.45 $\pm$ 0.031 $\times$ 0.32 $\pm$ 0.023, 0.22 $\pm 9.4$ \\
\enddata
\end{deluxetable*}
\begin{deluxetable*}{lcccc}
\label{tab:mass}
\tabletypesize{\scriptsize}
\tablecaption{Physical Parameters of Centrally Condensed Structure}
\tablewidth{0pt}
\tablehead{
\colhead{Data}  & \colhead{Temperature (K)}&\colhead{Gas mass (M$_\odot$) \tablenotemark{a}} & \colhead{Column density (10$^{23}$cm$^{-2}$)\tablenotemark{b}} & \colhead{Number density (cm$^{-3}$)\tablenotemark{b}} 
}
\startdata
ACA                & 20 & 1.1 $\pm$ 0.1  & 0.46 $\pm$ 0.01   & (2.7 $\pm$ 0.5) $\times$ 10$^6$\\
ACA                 & 40 & 0.47 $\pm$ 0.02  & 0.20 $\pm$ 0.01   & (1.0 $\pm$ 0.2) $\times$ 10$^6$\\
ACA + ALMA low resolution & 20 & 0.11 $\pm$ 0.01 & 3.2 $\pm$ 0.2  &(1.6 $\pm$ 0.6) $\times$ 10$^8$\\
ACA + ALMA low resolution& 40 & (4.7 $\pm$ 0.4) $\times 10^{-2}$ & 1.4 $\pm$ 0.1  & (6.8 $\pm$ 2.7) $\times$ 10$^7$\\
ALMA high resolution & 20 & (1.6 $\pm$ 0.1) $\times 10^{-2}$ & 6.9 $\pm$ 0.2  & (1.2 $\pm$ 0.3) $\times$ 10$^9$\\
ALMA high resolution  & 40 & (7.0 $\pm$ 0.4) $\times 10^{-3}$ & 3.0 $\pm$ 0.1 & (5.2 $\pm$ 0.1) $\times$ 10$^8$\\
\enddata
\tablenotetext{a}{The errors for the mass are obtained by propagating only the errors for the measured flux densities.}
\tablenotetext{b}{The errors for the column density and number density are obtained by propagating the errors for the measured flux densities and deconvolved sizes. } 
\end{deluxetable*}

\subsection{{\rm Dense Gas Tracers}} \label{subsec:C18ON2D+}
\begin{figure*}
    \centering
    \plotone{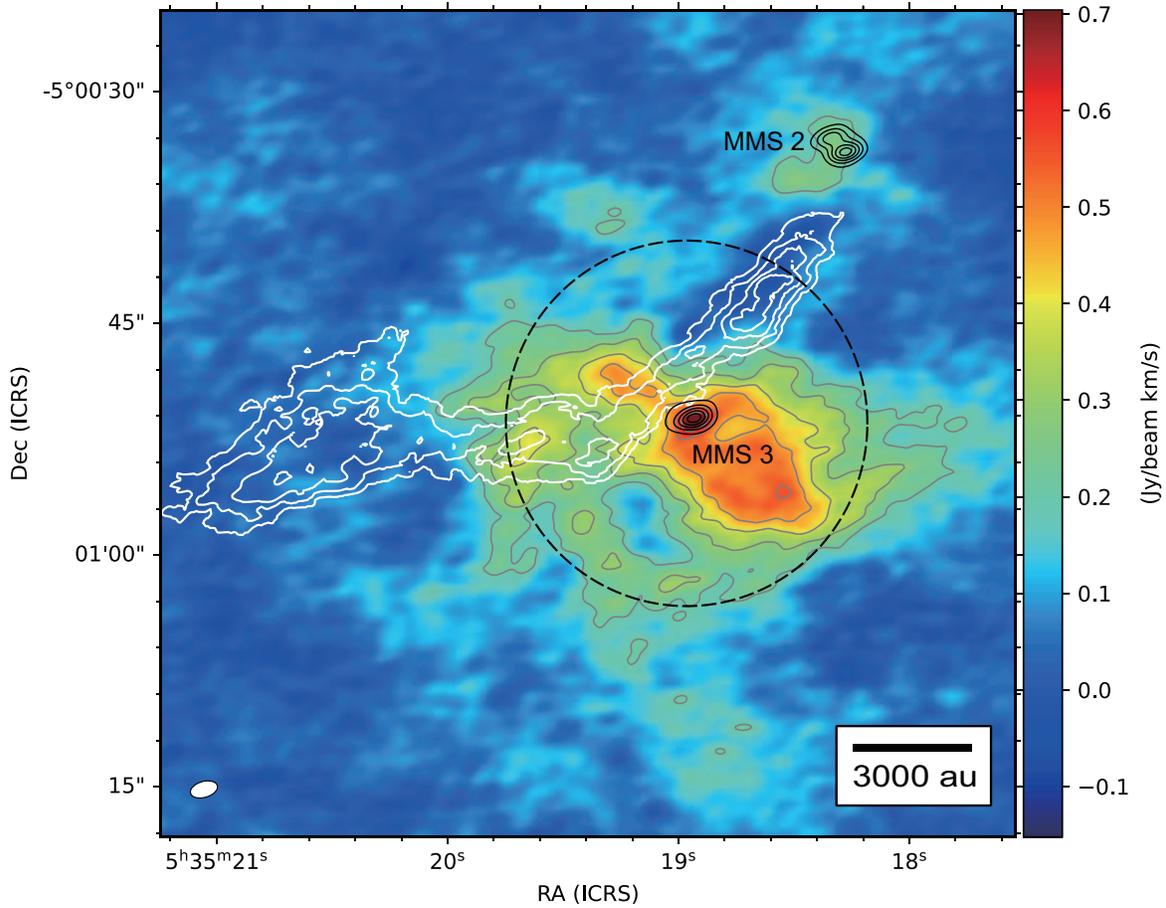}
    \caption{
C$^{18}$O ($J$ = 2--1) and N$_2$D$^+$($J$ = 3--2) emissions overlaid on the 1.3 mm continuum emission. 
The color shows the integrated intensity map for C$^{18}$O. The gray and white contours are the integrated intensity for C$^{18}$O and N$_2$D$^+$, respectively. The contours for C$^{18}$O start at 7$\sigma$ with an interval of 3$\sigma$ (1$\sigma$ = 29 mJy beam$^{-1}$\,km\,s$^{-1}$).
The contours for N$_2$D$^+$ correspond to 3, 4, 5, and 6$\sigma$ (1$\sigma$ = 34 mJy beam$^{-1}$\,km\,s$^{-1}$).
These emissions are integrated over the range $v_{\rm LSR}$ = 9--13.4 \,km\,s$^{-1}$.
The black contours correspond to the 1.3 mm continuum emission (ACA + ALMA low resolution), in which the contours start at 10$\sigma$ with an interval of 5$\sigma$ (1$\sigma$ = 0.46 mJy beam$^{-1}$). 
The ellipse in the bottom left corner is the beam size. 
The spatial scale is indicated in the bottom right corner.
The ALMA primary beam size is denoted by a black dashed circle. 
}
\label{fig:C18ON2D+}  
\end{figure*}

\begin{figure*}
    \centering
    \plotone{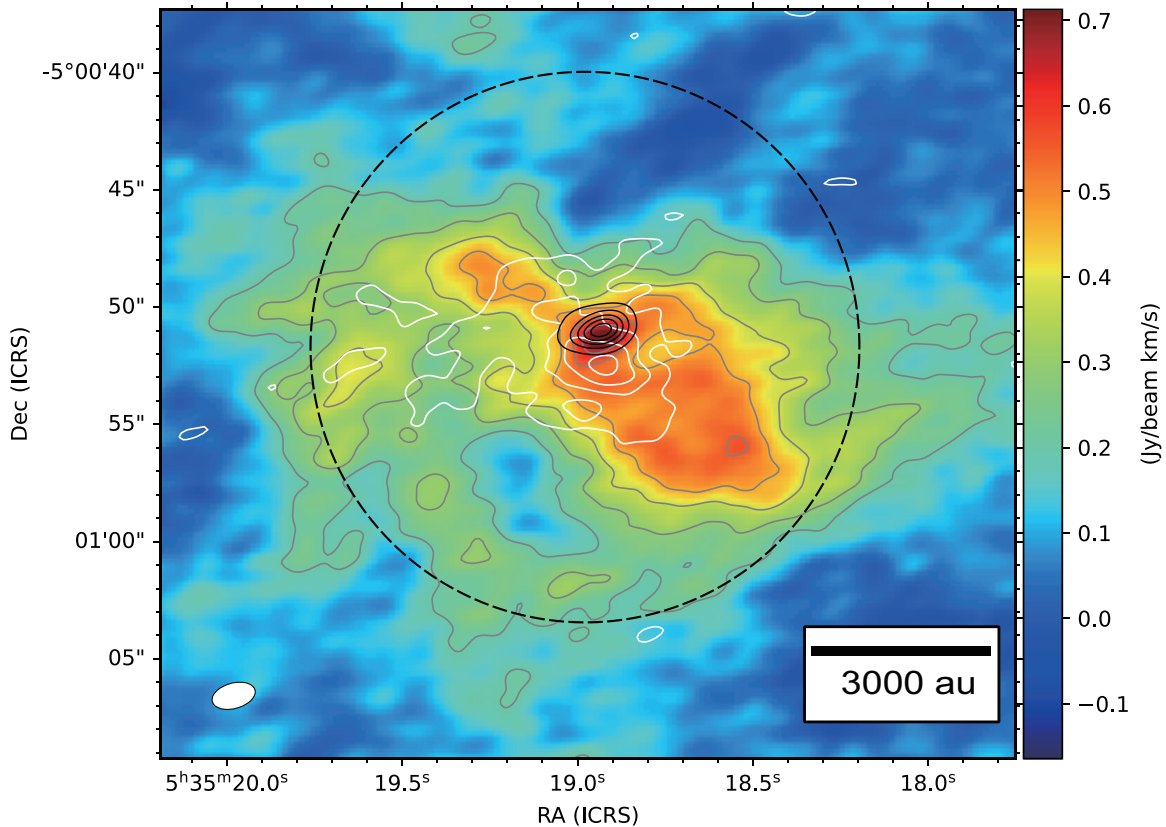}
    \caption{
C$^{18}$O ($J$ = 2--1) and DCN ($J$ = 3--2) emissions overlaid on the 1.3 mm continuum emission. 
The gray and black contours are the same as in Figure~\ref{fig:C18ON2D+}.
The white contours are the integrated intensity of DCN. 
The contours correspond to 3, 6, 9, and 12$\sigma$ (1$\sigma$ = 23 mJy beam$^{-1}$\,km\,s$^{-1}$).
These emissions are integrated over the range $v_{\rm LSR}$ = 6.2--15.7\,km\,s$^{-1}$.
The spatial scale is indicated in the bottom right corner.
The ellipse in the bottom left corner is the beam size.
The ALMA primary beam size is denoted by a black dashed circle. 
}
\label{fig:C18ODCN}  
\end{figure*}

\begin{figure*}
\begin{center}
\plotone{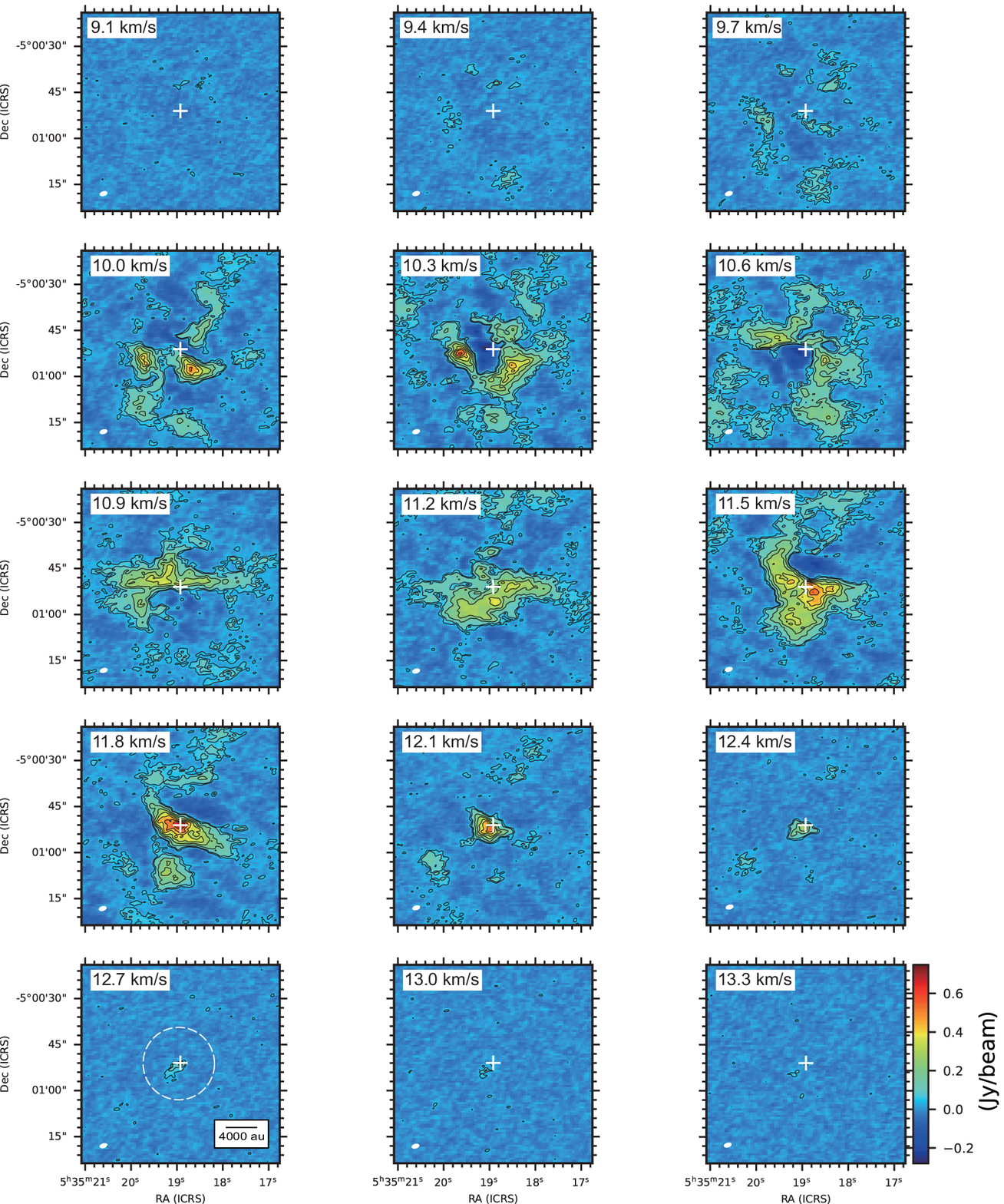}
\caption{
Channel maps for C$^{18}$O ($J$ = 2--1). 
In each panel, contour levels of 3, 5, 10, 15, 20, 25$\sigma$ (1$\sigma$= 16 mJy\,beam$^{-1}$) are plotted. 
The white symbol `+' corresponds to the peak position of the 1.3 mm continuum emission. 
The central velocity is shown in the upper left corner of each panel.
The ellipse in the bottom left corner of each panel is the beam size. The spatial scale and the ALMA primary beam (white dashed circle) are denoted in the bottom left panel. 
}
\label{fig:c18ochan} 
\end{center}
\end{figure*}

Figure~\ref{fig:C18ON2D+} shows the integrated intensity maps for C$^{18}$O ($J$ = 2--1) and N$_2$D$^+$($J$ = 3--2) emissions overlaid with the 1.3 mm dust thermal emission.
We detected a centrally condensed C$^{18}$O ($J$ = 2--1) emission structure around  the 1.3 mm continuum emission, presumably tracing the dense gas envelope
\footnote{In this paper, we define the term `envelope' as the circumstellar material surrounding  the 1.3 mm compact continuum emission traced by the dense gas tracers such as C$^{18}$O and DCN. We use the term `core'  to represent a large self-gravitating object, which corresponds to a molecular cloud core, as used in previous studies.} around MMS 3.
The long-axis length of the C$^{18}$O emission estimated by 2D Gaussian fitting is $\sim$6800\,au and it is elongated from the northeast to the southwest directions with P.A. = 75$^\circ$.

The N$_2$D$^+$ ($J$ = 3--2) emission distribution around MMS 3 
extends from the northwest to the southeast and is elongated to the east,  which is consistent with large-scale filaments in the OMC-3 region observed in the millimeter and submillimeter continuum emission \citep{chini1997,Johnstone_1999}.
The velocity width of the N$_2$D$^+$ emission is relatively 
narrow ($\delta v_{\rm FWHM} \sim$0.89 km s$^{-1}$). 

We detected a DCN ($J$ = 3--2) line in the spectral window of SiO ($J$ = 5--4) and produced the image presented in Figure~\ref{fig:C18ODCN}. 
The DCN emission shows a centrally condensed structure and seems to roughly overlap with the C$^{18}$O ($J$ = 2--1) emission.
The long-axis length of the DCN emission estimated by 2D Gaussian fitting is $\sim$2000\,au, which is more compact than that of the C$^{18}$O emission. 
The local peak of the DCN emission is shifted ${\sim}1''.3$ to the south with respect to the peak position of the 1.3 mm continuum and C$^{18}$O emissions.  
A temperature in the range of 10 to 80 K is  required for the main formation path of DCN molecules \citep{tuener01,Salinas2017}.
Hence it is natural that DCN traces the dense gas envelope except for the region very close to the protostar where the gas temperature should be as high as T$\ga$100 K.

Figure~\ref{fig:c18ochan} shows the channel maps for C$^{18}$O ($J$ = 2--1). 
In the velocity range of $v_{\rm LSR}$ = 9.7 to 10.3 \,km\,s$^{-1}$, 
the C$^{18}$O emission shows extended complex structures.
We can see two emission peaks located at $\sim$11$\arcsec$ southeast and $\sim$8$\arcsec$ southwest with respect to the 1.3 mm continuum peak.
Around a velocity of $v_{\rm LSR}$ = 10.6 \,km\,s$^{-1}$,  the C$^{18}$O emission is elongated from the southwest to the northeast across the 1.3 mm continuum emission peak.
The C$^{18}$O emission is distributed in the  north part toward  the 1.3 mm continuum peak around a velocity of  $v_{\rm LSR}$ = 10.9 \,km s$^{-1}$. 
The extended C$^{18}$O emission in the velocity range $v_{\rm LSR} = 10.6$--$10.9$\,km\,s$^{-1}$ shows a butterfly-like structure 
particularly in the northwest direction with respect to the 1.3 mm continuum peak. 
The distribution is presumably related to the outflow cavity (blue-shifted gas) observed in the CO emission (see \S\ref{subsec:outflow-distribution}). 
Hence, this structure likely traces the gas swept up by  the molecular outflow. 
The emission peak position in the channel map is significantly shifted from the north to the south across a systemic velocity of $v_{\rm LSR}$ = 11.2\,km\,s$^{-1}$.
In the range of  $v_{\rm LSR}$ = 11.5 to 12.1\,km s$^{-1}$,  the emission is condensed around the 1.3 mm continuum peak and shows an elongated structure along the northeast to the southwest. 
The elongation direction is almost perpendicular to the OMC-3 filament traced by N$_2$D$^+$ 
(Figure~\ref{fig:C18ON2D+}).
A compact C$^{18}$O emission associated with the 1.3 mm peak remains 
in the velocity range of $v_{\rm LSR}$ = 12.1 to 12.7\,km s$^{-1}$.

In this study, we adopted a systemic velocity for MMS 3 
of 11.2\,km\,s$^{-1}$, derived from 
the optically thin C$^{18}$O ($J$ = 2--1) and N$_2$D$^+$ ($J$ = 3--2) emissions.
This velocity corresponds to the central velocity
where we see a dip in these two lines. 
The dip is confirmed over all of the area of both lines because the ambient gas associated with the large-scale structure is resolved out due to the filtering effect of the interferometric observations.
The same value of the systemic velocity was confirmed in the optically thin H$^{13}$CO$^{+}$ ($J$ = 1--0)
and N$_2$H$^{+}$ ($J$ = 1--0) emissions in single-dish observations \citep{Ikeda07,Tatematsu08}.

Assuming optically thin C$^{18}$O emission and local thermodynamic equilibrium (LTE), the envelope column density ($N_{\mathrm H_2}$) is derived as
\begin{equation}
    N_{\rm H_2} = X_\mathrm{C^{18}O}^{-1}\left(\frac{3h}{8\pi^3S\mu^2}\right) \left(\frac{kT_\mathrm{ex}}{hB} + \frac{1}{3} \right) 
    \, {\rm exp}\left( \frac{E_\mathrm{u}}{kT_{\rm ex}} \right) \int T_{\rm B} dv,
    \label{eq;column}
\end{equation}   
where $X_\mathrm{C^{18}O}$, $S$, $\mu$, $T_{\rm ex}$, $E_u$, and $T_B$
are the C$^{18}$O to H$_2$ abundance ratio, the line strength, the relevant dipole moment, the excitation temperature, the energy of upper level above ground, and the brightness temperature in units of K, respectively \citep[]{Cabrit92, Magnum15,Feddersen20}.
Here, we adopted $X_{\rm C^{18}O}$ of $1.7 \times 10^{-7}$ \citep{Frerking82}, 
$S\mu^2$ of 0.02 Debye$^2$, and $E_u/k$ of 15.8\,K.
We assumed that the excitation temperature is $T_{\rm ex}$=20 K.
Then, using the estimated $N_{\mathrm H_2}$, the envelope mass $M_{\mathrm H_2}$ can be estimated as
\begin{equation}
    M_{\mathrm H_2} = \mu_\mathrm{H_2} m_{\rm H}\, \Omega \, d^2 N_\mathrm{H_2}, 
    \label{eq;mout}
\end{equation}
where $\Omega$ is the total solid angle, $\mu_\mathrm{H_2}=2.8$ is a mean molecular weight \citep[][]{Kauffmann08}, and $d$ is the source distance. 
The estimated envelope column density and mass using the flux more than 7$\sigma$ are $N_\mathrm{H_2} = 1.3 \times 10^{22}$cm$^{-2}$ and $M_\mathrm{H_2} = 0.89$\,M$_\odot$, respectively.

\subsection{\rm Outflow and Jet Tracers}
\subsubsection{\rm Morphology}
\label{subsec:outflow-distribution}

\begin{figure*}
\begin{center}
\plotone{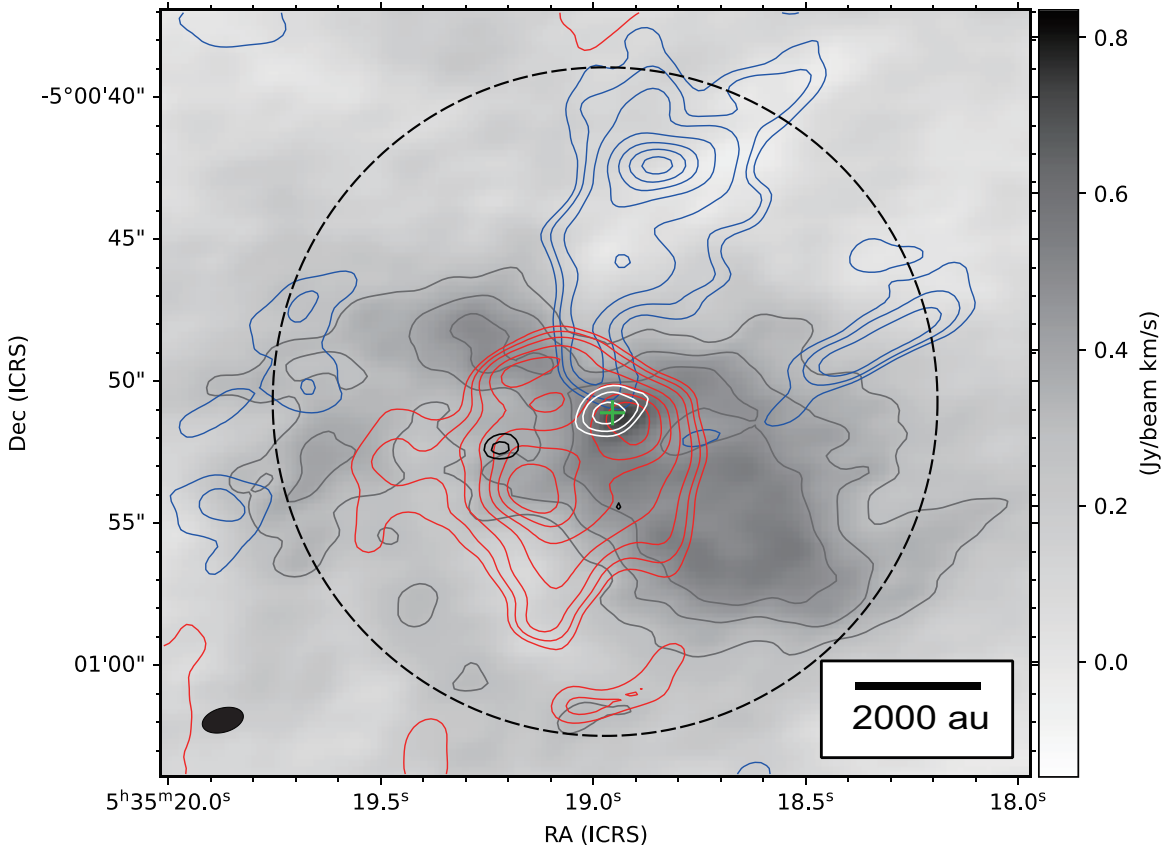}
 \caption{
Integrated intensity map for C$^{18}$O ($J$ = 2--1), CO ($J$ = 2--1), and SiO ($J$ = 5--4) emissions.
The grayscale and gray contours correspond to  C$^{18}$O integrated from 9  to 13.4 \,km\,s$^{-1}$.
The contour levels are  10, 13, and 16$\sigma$ (1$\sigma$ = 29 mJy beam$^{-1}$\,km\,s$^{-1}$).
The blue, red, and white contours are the integrated intensity of the CO emission, in which the emission is integrated from $v_{\rm LSR}=$ $-$4.8 to 11.1 \,km\,s$^{-1}$ (blue), from $v_{\rm LSR}=$11.2  to 33.2 \,km\,s$^{-1}$ (red), and from $v_{\rm LSR}=$33.2 to 41.2 \,km\,s$^{-1}$ (white), respectively.
The contour levels correspond to 5, 7, 10, 15, 20, 25, and 30 (1$\sigma$ = 150, 200, and 28 mJy\,beam$^{-1}$\,km\,s$^{-1}$). 
The SiO emission integrated from 11.2 to 15.2 \,km\,s$^{-1}$  is indicated by the black contours, in which the contour levels are  4$\sigma$ and 5$\sigma$ (1$\sigma$ = 2.9 mJy\,beam$^{-1}$\,km\,s$^{-1}$).
The spatial scale is indicated in the bottom right corner.
The black filled ellipse in the bottom left corner is the beam size. The ALMA primary beam size is denoted with a black dashed circle. 
}
\label{fig:COSiO} 
\end{center}
\end{figure*}

Figure~\ref{fig:COSiO} presents the integrated intensity maps 
for CO ($J$ = 2--1) and SiO ($J$ = 5--4). 
The CO emission was detected in the velocity range of 
$v_{\rm LSR}$ = $-$4.8 to 41.2\,km\,s$^{-1}$ 
with greater than 3$\sigma$ in the channel maps. 
The detected emission consists of two components.
The first is in the low- and mid-velocity ranges of 
$-$16.0 \,km\,s$^{-1} \leq v_{\rm LSR} - v_{\rm sys} \leq 22$ \,km\,s$^{-1}$, which traces the outflow cavity in both the blue- and red-shifted components.
The projected size of the blue- and red-shifted component is 5800 au and 4600 au, respectively. 
The position angle of the CO outflow cavity is $\sim$130\,$^\circ$, and  perpendicular to the long axis of the dense envelope observed in the C$^{18}$O emission. 
The second component is the high-velocity component, 
which is detected in the velocity range of 22 to 30 \,km\,s$^{-1}$  
with respect to the system velocity (i.e., red-shifted). 
The detected emission shows a compact structure associated with the 1.3 mm continuum peak, which may correspond to a recently ejected jet. 
In our observations, no blue-shifted high-velocity component was detected.

In addition to the CO emission, a compact 
SiO ($J$= 5--4) emission was marginally detected 
with the 5$\sigma$ emission peak in the integrated intensity map
($v_{\rm LSR} - v_{\rm sys} \simeq 1$--$4$\,km\,s$^{-1}$). 
The compact SiO emission was located 1800 au east 
from the 1.3 mm continuum emission peak position and gravitationally unbound.\footnote{
The infall $v_{\rm inf}$ and Keplerian $v_{\rm Kep}$ velocity are described as $v_{\rm inf} =(2GM/r)^{1/2}$ and $v_{\rm kep}=(GM/r)^{1/2}$, respectively. 
Assuming a protostellar mass of 0.1\,M$_\odot$ and the distance from the protostar of 1800\,au, the velocities are estimated to be an order of 0.1\,km\,s$^{-1}$ considering the inclination angle of 45$^\circ$.
The velocity of the detected SiO emission exceeds both the Keplerian and infall velocity.}.
The SiO emission is located in the outflow region traced by CO, while its velocity is not high. Thus, it is expected that the SiO emission originates from outgoing material, such as a low-velocity outflow.  Alternatively, this could be attributed to an outflow-envelope interaction.
More observations are necessary to identify the origin of the SiO emission. 

\subsubsection{\rm Outflow Parameters}
\label{sec:outflow-para}
To investigate the outflow properties, we derive outflow physical parameters from equations~(\ref{eq;column}) and (\ref{eq;mout}) assuming LTE  and an optically thin condition, 
in which X$_\mathrm{C^{18}O}$ in equation~(\ref{eq;column}) is replaced by X$_\mathrm{CO}$.
Here, we adopted the CO to H$_2$ abundance ratio $X_{\rm CO}$ of $10^{-4}$ \citep{Frerking82} and $E_u/k$ of 16.59\,K. 
We also assumed that the excitation temperature is $T_{\rm ex}$=20 K \citep[][]{Aso00}.
We only use the pixels above 3$\sigma$ in a given channel.
To obtain the total outflow mass, the blue-shifted and red-shifted CO emissions were integrated separately.
The blue-shifted $M_{\rm flow, b}$  and red-shifted $M_{\rm flow, r}$ outflow mass are estimated to be $M_{\rm flow, b} = 9.5 \times 10^{-4}\,\msun$ and  $M_{\rm flow, r}= 1.4 \times10^{-3}\,\msun$, respectively.

The intrinsic outflow velocity ($v_{\rm flow}$) and outflow length ($l_{\rm flow}$) can be calculated using the observed outflow velocity ($v_{\rm obs}$) and outflow length ($l_{\rm obs}$) as $v_{\rm flow} = v_{\rm obs} /$ cos\,$i$ and $l_{\rm flow} = l_{\rm obs} / $sin$\,i$, where $i$ is the inclination angle of the disk. 

We adopt an inclination angle of the disk $i$ = 45$^\circ$.
Considering the inclination angle, the blue-shifted and red-shifted maximum outflow velocities ($v_{\rm max, b}$ and $v_{\rm max, r}$) are estimated to be $v_{\rm max, b }=25\,\kms$ and $v_{\rm max, r}=35\,\kms$, respectively.

The outflow momentum $P_{\rm flow}$ and kinetic energy $E_{\rm flow}$ are defined as $P_{\rm flow}= M_{\rm flow} v_{\rm flow}$ and $E_{\rm flow}= M_{\rm flow} \, v_{\rm flow}^2/2$.
The momentum and energy of the blue-shifted outflow are $P_{\rm flow, b} = 8.2 \times 10^{-3}$ M$_\odot$ km s$^{-1}$ and
$E_{\rm flow, b} =1.2 \times 10^{42}$ erg, respectively, while those of the red-shifted outflow are $P_{\rm flow, r} = 1.0 \times 10^{-2}$ M$_\odot$ km s$^{-1}$ and $E_{\rm flow, r} = 3.1 \times 10^{42}$ erg, respectively.

The outflow time derivative quantities such as the mass loss rate, 
\begin {equation}
\dot{M}_{\rm flow} = \frac{M_{\rm flow}}{t_{\rm dyn}},
\end{equation}
the outflow momentum flux, 
\begin {equation}
F_{\rm flow} = \frac{P_{\rm flow}}{t_{\rm dyn}},
\end{equation}
and the outflow kinetic luminosity, 
\begin {equation}
L_{\rm kin} = \frac{E_{\rm flow}}{t_{\rm dyn}},
\end{equation}
are often used to evaluate the outflow activity, where $t_{\rm dyn}$ is the outflow dynamical timescale  estimated from the outflow length $l_{\rm flow}$ and the maximum velocity $v_{\rm max}$ as $t_{\rm dyn}= l_{\rm flow}/v_{\rm max}$.
The dynamical timescales for the blue-shifted and red-shifted outflow are estimated to be $t_{\rm dyn,b}= 1800$\,yr and $t_{\rm dyn,r}= 1000$\,yr, respectively.
The mass loss rate, outflow momentum flux and kinetic luminosity for the blue-shifted component can be estimated to be 
$\dot{M}_{\rm flow, b} = 5.3 \times 10^{-7}\,\msun$\,yr$^{-1}$,
$F_{\rm flow, b}=4.5 \times 10^{-6}\, \msun$\,km\,yr$^{-1}$, and 
$L_{\rm kin, b}=6.3 \times 10^{-3}\,L_\odot$, respectively.
Those derived for the red-shifted component are
$\dot{M}_{\rm flow, r} = 1.3 \times 10^{-6}\,\msun$\,yr$^{-1}$,
$F_{\rm flow, r}=9.9 \times 10^{-6}\, \msun$\,km\,yr$^{-1}$ and 
$L_{\rm kin, r}=2.7 \times 10^{-2}\,L_\odot$, respectively. 

\begin{figure*}
\begin{center}
\epsscale{1.1}
\plotone{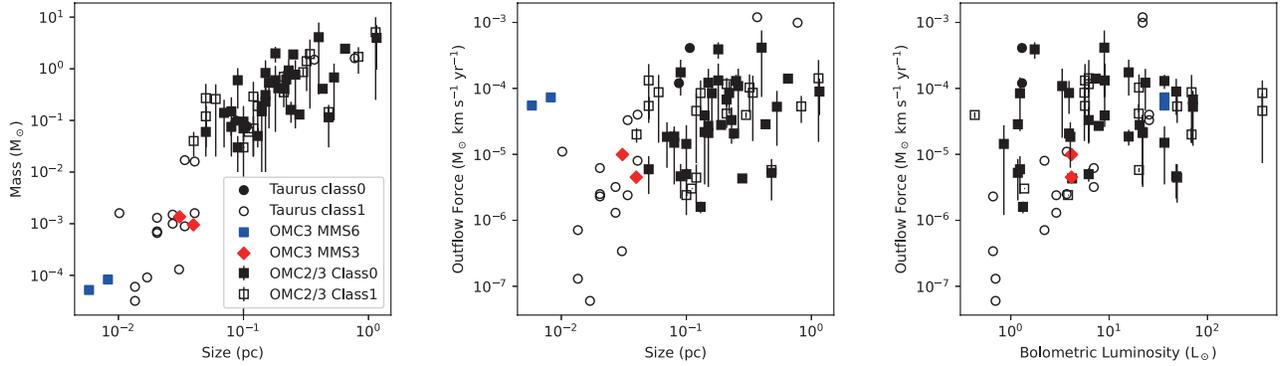}
 \caption{
Outflow size vs. outflow mass (left), outflow size vs. outflow force (middle), and bolometric luminosity vs. outflow force (right).
The red filled diamonds denote the outflow associated with OMC-3/MMS 3. 
The blue filled squares denote the outflow associated with OMC-3/MMS 6 \citep[][]{TakahashiandHo2012}.
The squares and circles represent the outflows observed in the OMC-2/3  \citep{Feddersen20} and the Taurus \citep{Hogerheijde98} star-forming regions, respectively. 
Class 0 and I objects are shown by filled and open symbols, respectively. 
The bolometric luminosity of the OMC-2/3 objects is taken from \citet{Furlan2016}.
Outflow parameters in OMC-2/3 estimated by \citet{Feddersen20} plotted here are consistent with previous several studies within factor of $\sim$3 on average \citep[i.e.,][]{Takahashi08,Tanabe19}.
}
\label{fig:outflow-para} 
\end{center}
\end{figure*}

To understand the evolutionary stage and the status of the outflow in MMS 3, in Figure~\ref{fig:outflow-para}, we compare the outflow size, mass, force (or momentum flux), and bolometric luminosity of MMS 3 with those of other outflows associated with Class 0 and Class I sources in the OMC-2/3 \citep[][]{Takahashi08,TakahashiandHo2012,Furlan2016,Tanabe19,Feddersen20} and Taurus star-forming regions \citep[][]{Hogerheijde98}.
In the figure, compared to other developed outflow associated with most of the intermediate-mass protostellar sources in the OMC-2/3 region, the outflow detected in MMS 3 is less massive and smaller.
Rather, the outflow in MMS 3 shares a similar nature to low-mass protostellar outflows detected in the Taurus star-forming region \citep[c.f.][]{Hogerheijde98}.

In our observations, we detected an outflow in MMS 3, which was not identified in the previous molecular outflow survey in this region \citep[][]{Aso00,Williams03,Takahashi08,Tanabe19,Feddersen20}. 
This implies that the outflow in MMS 3 is compact and significantly faint, as it would not be detected in a survey type observation because of the low sensitivity and low angular resolution. It should be noted that the angular resolution of this observation ($\sim1''$) is better than those of past studies (e.g., $\sim22''$ for \citealt{Takahashi08} and \citealt{Tanabe19} and  $\sim10''$  for \citealt{Feddersen20}).
Also, the sensitivity of this study is about seven times better than in previous studies. 

\section{Evolutionary Stage of MMS 3}
\label{sec:discussion}
\subsection{Observational Characteristics of MMS 3}

The target of our study, MMS 3, has been recognized as a  mysterious object in past studies because various observations of MMS 3 have exhibited both prestellar and protostellar natures. 
The peak (sub)millimeter flux for MMS 3 is  3--20 times weaker than other protostellar sources  observed in the OMC-3 region, such as MMS 2 (SMM 3), MMS 5 (SMM 6), MMS 6 (SMM 7), and MMS 7 (SMM 11), while it is comparable  to prestellar sources such as MMS 4 (SMM 5) and SMM 12  \citep[see Figure~3 of][]{takahashi2013}.  
The number density of hydrogen molecule of MMS 3 within $D \sim$3000\,au is $n \sim$(4$\pm2) \times 10^6$\,cm$^{-3}$, which is at least one order of magnitude lower  than those of the protostellar sources in this region.
Thus, it has been considered that the relatively low density (or low condensation) of MMS 3 is consistent with neither outflow nor jet detection \citep{Reipurth99,Takahashi08,Tanabe19,Feddersen20} because MMS 3 seems to be  in the prestellar phase. 
Nevertheless, both near-infrared and X-ray sources were  detected within MMS 3, which is strong evidence of the existence of a protostar \citep[][]{Tsuboi2001,Furlan2016}. 
The spectral energy distribution measured in the near- to far-infrared wavelengths suggests that MMS 3 is a Class 0 source.
These observational characteristics have motivated us to study the evolutionary stage of MMS 3  through direct imaging with high angular resolution and high sensitivity observations.

In this study, the first detection of a centrally condensed compact structure with a size of $\sim$150\,au, which plausibly traces  a disk (Figure~\ref{fig:continuum}{\it c}), was made by our ALMA dust continuum observations with an angular resolution of ${\theta} {\sim}0''.2$. 
The N$_2$D$^+$ emission was also detected and traces a large-scale filament in the OMC-3 region. 
On the other hand, the C$^{18}$O emission traces a flattened envelope with a size of ${\sim}$6800\,au, which is elongated perpendicular to the large-scale filament traced by N$_2$D$^+$.
The spatial distribution of the C$^{18}$O emission is very different from that of the N$_2$D$^+$ emission that traces the filament. 
The spatial discrepancy between C$^{18}$O and N$_2$D$^+$  is considered to be realized when parent molecules such as H$_2$D$^+$ are destroyed by proton exchange with CO after the gas temperature exceeds the sublimation temperature of the CO molecules ($T\gtrsim$20\,K; \citealt{Jorgensen2004, Crapsi05, Salinas2017}). 
Thus, it is expected that  a mature and warm envelope with a temperature  of $\gtrsim$20\,K  already exists  in MMS 3.
We also detected a CO bipolar outflow (Figure~\ref{fig:COSiO}), in which cavity-like structures seem to be interacting with the C$^{18}$O flattened envelope. 
The direction of the outflow is perpendicular to the long axis of the envelope. 
The compact disk, flattened envelope, and molecular outflow detected in our  ALMA observations indicate that MMS 3 is not a prestellar source, but a protostellar source, consistent with the detection of infrared and X-ray sources in MMS 3.

We also investigated the outflow properties of MMS 3.
The comparisons of the outflow physical properties between MMS 3 and other sources in the OMC-2/3 region are presented in Figure~\ref{fig:outflow-para}. 
The masses and sizes of the outflows associated with MMS 3 and MMS 6 are significantly smaller than those detected by the survey observations in the OMC-2/3 region \citep{Takahashi08,Tanabe19,Feddersen20}. This indicates that the outflows detected in MMS 3 and MMS 6 are very young. 
The outflow dynamical timescale, which is an indicator for identifying the protostellar age, estimated for MMS 3 is~1000-1800\,yr.
In addition, the dynamical timescale for MMS 6 \citep{TakahashiandHo2012,Takahashi2019} is as short as $\lesssim$1000\,yr. 
Thus, the outflow dynamical timescale is slightly longer in MMS 3 than in MMS 6. 
Both a clear outflow and a powerful jet were detected in strong CO and SiO emissions for MMS 6 \citep{Takahashi2012,Takahashi2019,hsu20}. 
It should be noted that a powerful jet was also detected with strong CO and SiO emissions in MMS 5 \citep[][Matsushita et al. in prep.]{Matsushita2019}.
These sources (MMS 3, MMS 5 and MMS 6) are embedded in the same filamentary cloud \citep{chini1997,takahashi2013}, in which MMS 5 and MMS 6 is located in the southwest of MMS 3. Nevertheless, MMS 3 does not show a powerful jet, unlike MMS 5 and MMS 6.  
As shown in Figure~\ref{fig:outflow-para}, the outflow force in MMS 6 is one order magnitude larger than that of MMS 3, although the mass and size of outflows in MMS 6 are smaller than those in MMS 3. Thus, there is a difference in the outflow strength between MMS 3 and a jet associated source, MMS 6.
We noticed that the outflow parameters for MMS 3 are similar to those for Class I protostars observed in the Taurus star-forming region \citep{Hogerheijde98}. Thus, the outflow nature of MMS 3 implies that a low-mass protostar with a relatively low accretion rate is forming there. 

The marginal detection (5$\sigma$ detection) of a very compact SiO emission (Figure~\ref{fig:COSiO}) could be attributed to a jet in MMS 3. \cite{Matsushita2019} showed that the intensity ratio between CO and SiO emission is $\sim$47 at their peak. Assuming a similar intensity ratio for MMS 3, the SiO emission can be detected at $\sim$4$\sigma$ signal level. This implies that SiO detection should be limited around the peak position in the CO emission. Thus, the marginal SiO detection for MMS 3 may be natural considering the limited sensitivity of our observations. It should be noted that, alternatively, the SiO emission is possibly related to the interaction between the outflow and the envelope because the velocity of the SiO emission is not very high, as described in \S\ref{subsec:outflow-distribution}. 
High sensitivity SiO observations will be needed to confirm either scenario.

\subsection{Possible Evolutionary Scenarios}

In this subsection, we discuss the evolutionary status of MMS 3.
As described above, in addition to a faint bipolar outflow, we detected a very compact high-velocity component around the continuum peak position (Figure\ref{fig:COSiO}). 
Additionally, the number density of MMS 3 is comparably low compared with other prestellar sources found in the OMC-3 region.
These characteristics of MMS 3 resemble the very early phase of star formation just after protostar formation. 
In the collapsing (prestellar) cloud, the first core drives a wide-angle bipolar outflow before protostar formation \citep{machida08, tomida13}. 
Then, the central region of the first core collapses to form a protostar that drives a collimated jet. 
Thus, a small-sized jet is enclosed by a wide-angle large-scale outflow just after protostar formation, as seen in Figure~\ref{fig:COSiO}. 
At this epoch,  the remnant of the first core  remains around the protostar with a size of $\sim$10--100\,au, which is comparable to the size of the condensed compact structure seen in Figure\ref{fig:continuum}(c).
The lifetime of the first core is $\sim$100--1000\,yr during which only the wide-angle outflow appears \citep{Saigo2006,tomida13}. 
The outflow dynamical timescale of $t_{\rm dyn}\sim$1000--1800\,yr  (\S3.3.2) agrees  well with the first core lifetime.  
In addition, from theoretical studies \citep[e.g.,][]{masunaga00}, the luminosity of the low-mass protostar at this epoch is estimated to be $\sim$1--$10 L_\odot$, which is comparable to the bolometric luminosity of MMS 3.
Furthermore, a protostellar source at this epoch would be observed to be similar to a prestellar source because the protostar has just been born, and its surrounding environment should not be affected by the protostar. 
Thus, the very early star formation phase scenario seems to explain the observational properties of MMS 3 well. 
However, the velocity of the outflow ($\gtrsim$10$\kms$) observed in MMS 3 is larger than the theoretical prediction. 
Theoretical studies have shown that  the outflow velocity just after protostar formation is limited  to $<10\,\kms$ \citep[e.g.,][]{machida08}. 
Besides, as shown in \S\ref{subsec:C18ON2D+}, the C${^{18}}$O and DCN distributions imply that the central protostar is surrounded by a warm envelope with $T > 20$\,K. 
In addition, the peak shift of DCN (1$\farcs 3$ corresponding to $\sim$520 au) suggests the presence of high-temperature ($>$ 100 K) gas in the vicinity of the central star. 
These results seem to contradict the very early phase scenario.

An alternative scenario to explain the observed characteristics of MMS 3 is the low-mass star formation scenario. 
It is considered that intermediate-mass stars in the OMC-3 region preferentially form with high accretion rates, because of the high mass ejection rate, which is proportional to the mass accretion rate \citep{Takahashi08,TakahashiandHo2012,Matsushita2019}. 
On the other hand, the outflow parameters for MMS 3 are similar to those for low-mass stars observed in the low-mass star-forming region (Fiure~\ref{fig:outflow-para}). 
Thus, the mass accretion rate onto the protostar in MMS 3 is expected to be low. 
The mass accretion rate is determined by the condition of the prestellar phase. 
As described above, the number density within the central 3000\,au region is at least one order of magnitude smaller in MMS 3 than in other intermediate-mass Class 0 and I sources in the OMC-3 region \citep{takahashi2013}.
The relatively low mean number density corresponds to the less massive core within which a low-mass star with a relatively low mass accretion rate forms, instead of forming an intermediate-mass star with a relatively high mass accretion rate.
\citet{Matsushita2017} showed that a low mass accretion rate is realized, and a low-mass star forms when the natal cloud core is less dense (for details, see also \citealt{Machida2020})
\footnote{
\citet{Matsushita2017} pointed out that the accretion rate depends on the concentration of the star-forming cloud,  and a higher accretion rate is realized in a cloud with an initially higher central density (or higher central condensation). 
}.
Furthermore, the bolometric luminosity of MMS 3 is   $L_{\rm{bol}}$=4.2 or 3.6\,L$_{\odot}$, which is four to ten times lower than that of other protostellar sources in the OMC-3 region \citep{Furlan2016,Tobin2020}. 
During the main accretion phase, the bolometric luminosity is roughly proportional to the mass accretion rate.
Thus, the low bolometric luminosity also means that  the mass accretion rate is low
\footnote{ 
The relatively low luminosity of MMS 3 may be explained by episodic accretion.
The time varying accretion (or episodic accretion) temporally changes the bolometric luminosity. 
The bolometric luminosity becomes low in the low-accretion (or quiescent) phase. 
It should be noted that although the outflow intermittently appears with episodic accretion, we could not confirm any sign of time variability in the MMS 3 outflow (e.g., Figure~\ref{fig:outflow-para} right panel).
}.
Therefore, we expect that, in MMS 3,  a young protostar is growing with a low mass accretion rate.
As a result, a low-mass star would form in MMS 3, while intermediate-mass stars form in other bright millimeter sources such as MMS 5 and MMS 6 in this region. 

In summary, compared to other protostellar sources in the OMC-3 region, MMS 3 is peculiar because
the bolometric luminosity, number density, and outflow parameters, such as outflow force, are as low as those normally observed in low-mass protostellar sources. 
Also, no strong SiO emission was detected, in contrast to other Class 0 sources in OMC-3 \citep[][Takahashi et al. in prep.]{Matsushita2019,hsu20}.
To explain the characteristics of MMS 3, we proposed two possible scenarios: 
(1) a very early  star formation phase and (2) low-mass star formation with a low accretion rate.
For scenario (1), the protostar in MMS 3 is younger than the protostars in other protostellar sources and is in the very early phase just after protostar formation. 
However, the outflow velocity of $>10\,\kms$ cannot support this scenario.  
For scenario (2),  a low-mass star is currently forming with a low mass accretion rate  in MMS 3, while intermediate-mass stars with  relatively high mass accretion rates are forming in other sources. 
At the moment, the observational signatures do  not contradict scenario (2).  
The difference in observational characteristics between MMS 3 and other bright sources may be attributed to the initial condition of star formation and the surrounding environment of star forming cores, suggested by the previous SMA observations \citep[][]{takahashi2013}.

\section{Summary} \label{sec:summary}
We investigated the protostellar source MMS 3 in the OMC-3 region with 1.3 mm continuum, CO ($J$ = 2--1), C$^{18}$O ($J$ = 2--1), SiO ($J$ = 5--4),  N$_2$D$^+$ ($J$ = 3--2), and DCN ($J$ = 3--2) emissions and obtained the following results.
\begin{itemize}
\item 
With sub-arcsecond angular resolution, we detected compact 1.3 mm continuum sources with a size of $\sim$150 au and a mass of $\sim$10$^{-2}\,M_\odot$ for the first time, presumably tracing a compact dusty disk. 
The peak position of the 1.3 mm continuum emission corresponds to both the infrared (HOPS 91) and X-ray sources. 

\item 
A flattened envelope with a  size of $\sim$6800\,au was detected in the C$^{18}$O ($J$ = 2--1) emission.
A faint and compact bipolar outflow was also detected in the CO ($J$=2--1) emission for the first time.
The outflow direction is roughly perpendicular to  the major axis of the flattened envelope and disk detected in the  C$^{18}$O and 1.3 mm continuum emissions,  respectively. 
The outflow maximum gas velocity is $\sim$42\,km s$^{-1}$, assuming a disk inclination angle of $i$ = 45$^\circ$.
The SiO ($J$ = 5--4) emission is marginally detected at almost the same position as the CO outflow. 

\item 
Two possible scenarios were proposed to explain the evolutionary stage and status of a protostar in MMS 3. 
One is the low-mass Class 0 stage scenario, while the other is the very early phase scenario.
Although many observational characteristics can be explained by both scenarios, the high outflow velocity cannot support the latter. 
Comparing the outflow properties in MMS 3 with those in different star-forming regions, it is expected that a low-mass star is forming, for which the accretion rate in MMS 3 is expected to be lower than those of protostellar sources in the OMC-3 region. 
\end{itemize}

\acknowledgments
We thank the anonymous referee for providing us very helpful comments and suggestions.
This paper uses the following ALMA data: ADS/JAO. ALMA No. 2015.1.00341.S.  
ALMA is a partnership of ESO (representing its member states), NSF (USA) and NINS (Japan), together with NRC (Canada), $MOST$ and ASIAA (Taiwan), and KASI (Republic of Korea), in cooperation with the Republic of Chile. The Joint ALMA Observatory is operated by ESO, AUI/NRAO, and NAOJ.
K. Morii is very grateful for support from the SOKENDAI and NAOJ Chile Observatory (currently, NAOJ ALMA Project) while visiting a coauthor S. Takahashi through the SOKENDAI summer student program 2018.
This work was supported by JSPS KAKENHI grants JP17KK0096, JP17K05387, and 	JP17H06360.

\bibliography{reference}

\begin{thebibliography}{}
\expandafter\ifx\csname natexlab\endcsname\relax\def\natexlab#1{#1}\fi
\providecommand{\url}[1]{\href{#1}{#1}}
\providecommand{\dodoi}[1]{doi:~\href{http://doi.org/#1}{\nolinkurl{#1}}}
\providecommand{\doeprint}[1]{\href{http://ascl.net/#1}{\nolinkurl{http://ascl.net/#1}}}
\providecommand{\doarXiv}[1]{\href{https://arxiv.org/abs/#1}{\nolinkurl{https://arxiv.org/abs/#1}}}

\bibitem[{{Aso} {et~al.}(2000){Aso}, {Tatematsu}, {Sekimoto}, {Nakano},
  {Umemoto}, {Koyama}, \& {Yamamoto}}]{Aso00}
{Aso}, Y., {Tatematsu}, K., {Sekimoto}, Y., {et~al.} 2000, \apjs, 131, 465

\bibitem[{Bate(1998)}]{Bate1998}
Bate, M.~R. 1998, \apj, 508, L95

\bibitem[{Cabrit \& Bertout(1992)}]{Cabrit92}
Cabrit, S., \& Bertout, C. 1992, A\&A, 261, 274

\bibitem[{Chini {et~al.}(1997)Chini, Reipurth, Ward-Thompson, Bally, Nyman,
  Sievers, \& Billawala}]{chini1997}
Chini, R., Reipurth, B., Ward-Thompson, D., {et~al.} 1997, \apjl, 474, L135

\bibitem[{{Crapsi} {et~al.}(2005){Crapsi}, {Caselli}, {Walmsley}, {Myers},
  {Tafalla}, {Lee}, \& {Bourke}}]{Crapsi05}
{Crapsi}, A., {Caselli}, P., {Walmsley}, C.~M., {et~al.} 2005, \apj, 619, 379

\bibitem[{{Feddersen} {et~al.}(2020){Feddersen}, {Arce}, {Kong}, {Suri},
  {S{\'a}nchez-Monge}, {Ossenkopf-Okada}, {Dunham}, {Nakamura}, {Shimajiri}, \&
  {Bally}}]{Feddersen20}
{Feddersen}, J.~R., {Arce}, H.~G., {Kong}, S., {et~al.} 2020, \apj, 896, 11

\bibitem[{{Frerking} {et~al.}(1982){Frerking}, {Langer}, \&
  {Wilson}}]{Frerking82}
{Frerking}, M.~A., {Langer}, W.~D., \& {Wilson}, R.~W. 1982, \apj, 262, 590

\bibitem[{Furlan {et~al.}(2016)Furlan, Fischer, Ali, Stutz, Stanke, Tobin,
  Megeath, Osorio, Hartmann, Calvet, Poteet, Booker, Manoj, Watson, \&
  Allen}]{Furlan2016}
Furlan, E., Fischer, W.~J., Ali, B., {et~al.} 2016, \apjs, 224, 5

\bibitem[{{Hogerheijde} {et~al.}(1998){Hogerheijde}, {van Dishoeck}, {Blake},
  \& {van Langevelde}}]{Hogerheijde98}
{Hogerheijde}, M.~R., {van Dishoeck}, E.~F., {Blake}, G.~A., \& {van
  Langevelde}, H.~J. 1998, \apj, 502, 315

\bibitem[{{Hsu} {et~al.}(2020){Hsu}, {Liu}, {Liu}, {Sahu}, {Hirano}, {Lee},
  {Tatematsu}, {Kim}, {Juvela}, {Sanhueza}, {He}, {Johnstone}, {Qin},
  {Bronfman}, {Chen}, {Dutta}, {Eden}, {Jhan}, {Kim}, {Kuan}, {Kwon}, {Lee},
  {Lee}, {Moraghan}, {Rawlings}, {Shang}, {Soam}, {Thompson}, {Traficante},
  {Wu}, {Yang}, \& {Zhang}}]{hsu20}
{Hsu}, S.-Y., {Liu}, S.-Y., {Liu}, T., {et~al.} 2020, \apj, 898, 107

\bibitem[{{Ikeda} {et~al.}(2007){Ikeda}, {Sunada}, \& {Kitamura}}]{Ikeda07}
{Ikeda}, N., {Sunada}, K., \& {Kitamura}, Y. 2007, \apj, 665, 1194

\bibitem[{Johnstone \& Bally(1999)}]{Johnstone_1999}
Johnstone, D., \& Bally, J. 1999, The Astrophysical Journal, 510, L49

\bibitem[{J{\o}rgensen {et~al.}(2004)J{\o}rgensen, Schöier, \& van
  Dishoeck}]{Jorgensen2004}
J{\o}rgensen, J.~K., Schöier, F.~L., \& van Dishoeck, E.~F. 2004, A\&A, 416,
  603

\bibitem[{{Kauffmann} {et~al.}(2008){Kauffmann}, {Bertoldi}, {Bourke}, {Evans},
  \& {Lee}}]{Kauffmann08}
{Kauffmann}, J., {Bertoldi}, F., {Bourke}, T.~L., {Evans}, N.~J., I., \& {Lee},
  C.~W. 2008, \aap, 487, 993

\bibitem[{Larson(1969)}]{larson1969}
Larson, R.~B. 1969, \mnras, 145, 271

\bibitem[{{Machida} \& {Hosokawa}(2020)}]{Machida2020}
{Machida}, M.~N., \& {Hosokawa}, T. 2020, \mnras

\bibitem[{Machida {et~al.}(2010)Machida, Inutsuka, \& Matsumoto}]{Machida2010}
Machida, M.~N., Inutsuka, S., \& Matsumoto, T. 2010, \apj, 724, 1006

\bibitem[{{Machida} {et~al.}(2008){Machida}, {Inutsuka}, \&
  {Matsumoto}}]{machida08}
{Machida}, M.~N., {Inutsuka}, S.-i., \& {Matsumoto}, T. 2008, \apj, 676, 1088

\bibitem[{Machida \& Matsumoto(2012)}]{Machida2012}
Machida, M.~N., \& Matsumoto, T. 2012, \mnras, 421, 588

\bibitem[{Mangum \& Shirley(2015)}]{Magnum15}
Mangum, J.~G., \& Shirley, Y.~L. 2015, PASP, 127, 266

\bibitem[{{Masunaga} \& {Inutsuka}(2000)}]{masunaga00}
{Masunaga}, H., \& {Inutsuka}, S. 2000, \apj, 531, 350

\bibitem[{{Mathis} {et~al.}(1977){Mathis}, {Rumpl}, \& {Nordsieck}}]{Mathis77}
{Mathis}, J.~S., {Rumpl}, W., \& {Nordsieck}, K.~H. 1977, \apj, 217, 425

\bibitem[{Matsushita {et~al.}(2017)Matsushita, Machida, Sakurai, \&
  Hosokawa}]{Matsushita2017}
Matsushita, Y., Machida, M.~N., Sakurai, Y., \& Hosokawa, T. 2017, \mnras, 470,
  1026

\bibitem[{Matsushita {et~al.}(2019)Matsushita, Takahashi, Machida, \&
  Tomisaka}]{Matsushita2019}
Matsushita, Y., Takahashi, S., Machida, M.~N., \& Tomisaka, K. 2019, \apj, 871,
  221

\bibitem[{McMullin {et~al.}(2007)McMullin, Waters, Schiebel, \&
  Young}]{McMullin07}
McMullin, J.~P., Waters, B., Schiebel, D., \& Young, W.;~Golap, K. 2007, ASPC,
  376, 127

\bibitem[{Megeath {et~al.}(2012)Megeath, Gutermuth, Muzerolle, Kryukova,
  Flaherty, Hora, Allen, Hartmann, Myers, Pipher, Stauffer, Young, \&
  Fazio}]{Megeath2012}
Megeath, S.~T., Gutermuth, R., Muzerolle, J., {et~al.} 2012, \aj, 144, 192

\bibitem[{Nielbock {et~al.}(2003)Nielbock, Chini, \& Müller}]{Nielbock2003}
Nielbock, M., Chini, R., \& Müller, S. A.~H. 2003, A\&A, 408, 245

\bibitem[{Ossenkopf \& Henning(1994)}]{OssenkopfHenning1994}
Ossenkopf, V., \& Henning, T. 1994, A\&A, 291, 943

\bibitem[{{Reipurth} {et~al.}(1999){Reipurth}, {Rodr{\'\i}guez}, \&
  {Chini}}]{Reipurth99}
{Reipurth}, B., {Rodr{\'\i}guez}, L.~F., \& {Chini}, R. 1999, \aj, 118, 983

\bibitem[{Saigo \& Tomisaka(2006)}]{Saigo2006}
Saigo, K., \& Tomisaka, K. 2006, \apj, 645, 381

\bibitem[{Salinas {et~al.}(2017)Salinas, Hogerheijde, Mathews, Öberg, Qi,
  Williams, \& Wilner}]{Salinas2017}
Salinas, V.~N., Hogerheijde, M.~R., Mathews, G.~S., {et~al.} 2017, A\&A, 606,
  A125

\bibitem[{Takahashi {et~al.}(2013)Takahashi, Ho, Teixeira, Zapata, \&
  Su}]{takahashi2013}
Takahashi, S., Ho, P.~T., Teixeira, P.~S., Zapata, L.~A., \& Su, Y.-N. 2013,
  \apj, 763, 57

\bibitem[{Takahashi \& Ho(2012)}]{TakahashiandHo2012}
Takahashi, S., \& Ho, P. T.~P. 2012, \apj, 745, L10

\bibitem[{{Takahashi} {et~al.}(2009){Takahashi}, {Ho}, {Tang}, {Kawabe}, \&
  {Saito}}]{Takahashi09}
{Takahashi}, S., {Ho}, P. T.~P., {Tang}, Y.-W., {Kawabe}, R., \& {Saito}, M.
  2009, \apj, 704, 1459

\bibitem[{Takahashi {et~al.}(2019)Takahashi, Machida, Tomisaka, Ho, Fomalont,
  Nakanishi, \& Girart}]{Takahashi2019}
Takahashi, S., Machida, M.~N., Tomisaka, K., {et~al.} 2019, \apj, 872, 70

\bibitem[{Takahashi {et~al.}(2012)Takahashi, Saigo, Ho, \&
  Tomida}]{Takahashi2012}
Takahashi, S., Saigo, K., Ho, P. T.~P., \& Tomida, K. 2012, \apj, 752, 10

\bibitem[{{Takahashi} {et~al.}(2008){Takahashi}, {Saito}, {Ohashi}, {Kusakabe},
  {Takakuwa}, {Shimajiri}, {Tamura}, \& {Kawabe}}]{Takahashi08}
{Takahashi}, S., {Saito}, M., {Ohashi}, N., {et~al.} 2008, \apj, 688, 344

\bibitem[{{Tanabe} {et~al.}(2019){Tanabe}, {Nakamura}, {Tsukagoshi},
  {Shimajiri}, {Ishii}, {Kawabe}, {Feddersen}, {Kong}, {Arce}, {Bally},
  {Carpenter}, \& {Momose}}]{Tanabe19}
{Tanabe}, Y., {Nakamura}, F., {Tsukagoshi}, T., {et~al.} 2019, \pasj, 71, S8

\bibitem[{{Tatematsu} {et~al.}(2008){Tatematsu}, {Kandori}, {Umemoto}, \&
  {Sekimoto}}]{Tatematsu08}
{Tatematsu}, K., {Kandori}, R., {Umemoto}, T., \& {Sekimoto}, Y. 2008, \pasj,
  60, 407

\bibitem[{Tobin {et~al.}(2020)Tobin, Sheehan, Megeath,
  D{\'{\i}}az-Rodr{\'{\i}}guez, Offner, Murillo, van~'t Hoff, van Dishoeck,
  Osorio, Anglada, Furlan, Stutz, Reynolds, Karnath, Fischer, Persson, Looney,
  Li, Stephens, Chandler, Cox, Dunham, Tychoniec, Kama, Kratter, Kounkel,
  Mazur, Maud, Patel, Perez, Sadavoy, Segura-Cox, Sharma, Stephenson, Watson,
  \& Wyrowski}]{Tobin2020}
Tobin, J.~J., Sheehan, P.~D., Megeath, S.~T., {et~al.} 2020, \apj, 890, 130

\bibitem[{{Tomida} {et~al.}(2013){Tomida}, {Tomisaka}, {Matsumoto}, {Hori},
  {Okuzumi}, {Machida}, \& {Saigo}}]{tomida13}
{Tomida}, K., {Tomisaka}, K., {Matsumoto}, T., {et~al.} 2013, \apj, 763, 6

\bibitem[{Tomisaka(2002)}]{Tomisaka2002}
Tomisaka, K. 2002, \apj, 575, 306

\bibitem[{Tsuboi {et~al.}(2001)Tsuboi, Koyama, Hamaguchi, Tatematsu, Sekimoto,
  Bally, \& Reipurth}]{Tsuboi2001}
Tsuboi, Y., Koyama, K., Hamaguchi, K., {et~al.} 2001, \apj, 554, 734

\bibitem[{{Turner}(2001)}]{tuener01}
{Turner}, B.~E. 2001, \apjs, 136, 579

\bibitem[{{Williams} {et~al.}(2003){Williams}, {Plambeck}, \&
  {Heyer}}]{Williams03}
{Williams}, J.~P., {Plambeck}, R.~L., \& {Heyer}, M.~H. 2003, \apj, 591, 1025

\end{thebibliography}

\bibliographystyle{aasjournal}

\end{document}